\documentclass{PoS}
\usepackage{wrapfig} 
\usepackage{graphicx}
\usepackage{subfigure}

\title{PDF sensitivity studies using electroweak processes at LHCb }

\ShortTitle{PDF sensitivity studies at LHCb}

\author{\speaker{Francesco De Lorenzi}\thanks{The author would like to acknowledge the support of Irish Research Council for Science, Engineering and Technology.}\\
       University College Dublin\\
        E-mail: \email{francesco.de.lorenzi@cern.ch}}

\abstract{We describe parton density function sensitivity studies, using muon final
states produced through the Drell-Yan process via W, Z or $\gamma*$ down
to a $Q^2$ of $10$ $GeV^2$. This makes use of LHCb's unique ability to trigger on
low transverse momentum objects. Due to the forward acceptance of LHCb,
x values down to $2x10^{-6}$ can be probed, where with just $100$ $pb^{-1}$  of data,
the gluon PDF can improved up to $70\%$.}

\FullConference{XVIII International Workshop on Deep-Inelastic Scattering and Related Subjects, DIS 2010\\
		April 19-23, 2010\\
		Firenze, Italy}

\begin{document}

\section{Introduction}

LHCb \cite{tdr} is a detector designed to study b-physics so its pseudo-rapidity coverage extends in the forward region where B mesons are dominantly produced ($1.9 < \eta < 4.9$). Part of that range is in common with ATLAS and CMS which have a coverage of $|\eta|<2.5$; the rest is unique to LHCb.
During the first year of data taking,  LHC will run at the center of mass energy of $7$ $TeV$ and an integrated luminosity of $1$ $fb^{-1}$ will be delivered to LHCb in the first year of data taking. A unique feature of LHCb is its ability to trigger and reconstruct very low transverse momentum muons
($p_T>1 GeV$). Besides b-physics goals LHCb is able to perform precise measurement of electroweak channels with muon final states at high rapidities. We will show that these measurements can probe parton distribution functions for different values of Bjorken-x.
\section{LHCb parton kinematics}
\begin{wrapfigure}{r}{0.38\columnwidth}
  \vspace{-40pt}
  \begin{center}
    \includegraphics[width=0.38\columnwidth]{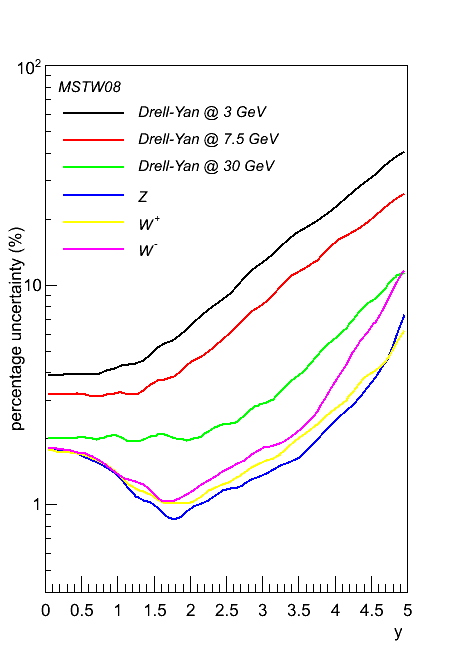}
    \vspace{-30pt}
    \caption{Percentage uncertainty on production cross-section of electroweak boson as function of rapidity}\label{uncert}
 \end{center}
  \vspace{-10pt}
\end{wrapfigure}
The current knowledge of the PDFs arises from fitting procedures to HERA , fixed target experiments and Tevatron data (or a subset of those).
 The foreseen electroweak measurements at LHCb include the analysis of the $Z$ and $W^{\pm}$ production and the Drell-Yan pair production of muons in four different mass bins ($2.5-5~GeV$, $5-10~GeV$, $10-20~GeV$ and $20-40~GeV$ )\cite{james}.
Production at high rapidity involves the interaction of one valence quark carrying a large fraction of proton momentum (high-x) and a sea quark with small fraction of momentum (very small x); according to the relation that link $x$ and $Q^2$
$$x_{1,2}=\frac{Q}{\sqrt{s}}e^{\pm y}$$ 
where $\sqrt{s}$ is the energy in the center of mass of the collisions. %
For the Drell-Yan production with $Q$ in the region $2.5-5~GeV$ the lowest value of $x$ reachable is $\sim 2x10^{-6}$. Hence due to its forward rapidity coverage and the ability to reconstruct low invariant mass muon pair systems, LHCb will cover an un-explored region of kinematical space where Parton Distribution Functions are known with less precision. 
This aspect can be also deduced looking at the theoretical uncertainties on the rapidity distribution for $W^{\pm}$, and $Z/\gamma{*}$; The largest contribution to these uncertainties arises from PDFs uncertainties. In figure \ref{uncert} the percentage uncertainties on the production cross-sections for  $W^{\pm}$, $Z$ and Drell-Yan process at three different mass are reported: the uncertainties due to PDF become large at high rapidity for all the channels studied (corresponding to rapidity LHCb acceptance) and for the Drell-Yan pair production, the smaller the mass, the larger the uncertainty.

\section{The fitting algorithm}
 PDFs are obtained via global fits to experimental data by different working groups, and their results have some model-dependence that may differ between each other. Theoretical predictions for the PDF sets MSTW08\cite{mstw}, CTEQ66\cite{cteq}, Alekhin\cite{alekh} and NNPDF2.0\cite{nnpdf} are used in this analysis.
 The first three groups perform a global fit based on a Hessian method while the fourth uses a neural network approach based on Montecarlo replicas. We now briefly describe our method to constrain the PDFs where the PDFs set have been obtained using these two approaches.

\subsection{PDFs built using a Hessian method}  \label{eigenv}
The Hessian method is used by the groups MSTW, CTEQ and Alekhin. A polynomial parametrization of the PDFs such as 
$xf(x,Q_0)= (1-x)^{\eta}(1+\varepsilon x^{0.5}+\gamma x )x^{\delta}$
is chosen. %
The global minimum of their fits to the global data produces a central value with the uncertainties given by a $N\times N$ covariance matrix. Diagonalizing the matrix gives a set of $N$ independent variables. In this way a representation of the PDF in a space where uncertainty can be treated as independent with a set of $N$ eigenvectors is provided.

A given physical observable, for example the rapidity distribution, $f(y)$, can be described as a linear combination of
 the observable obtained using the PDF best-fit, $f_0(y)$, and the observable $f_i(y)$ obtained by moving the $i^{th}$ eigenvector by 1-sigma.
\begin{equation}\label{theory}
f(y)=f_{0}(y)+\sum^N_{i=1}\left[f_{0}(y)-f_i(y)\right]\lambda_i
\end{equation}
The one sigma bound is mapped out by sampling $\lambda_i$ from a unit Gaussian with mean zero; this function also provides a form with which the data can be fitted.
Given binned data with $N_j$ events in the $j^{th}$ bin with error $\delta_j$, the normalization and the shape of the distribution can be fitted by minimizing
$$
\chi^2 = \sum_{j}^{\#bins}\left[\frac{N_{j}-\lambda_0\left(f_0+\sum_{i=1}^{N} \lambda_i(f_0-f_i) \right)}{\delta_{j}}\right]^2+\sum_{\lambda>0}\lambda_i^2
$$
where $\lambda_0$ is the overall normalization or luminosity and $\lambda_i$ are the fitted eigenvalues, constrained by the second term to follow a Gaussian probability.
Before the fit the eigenvalues by definition of the model have an uncertainty equal to 1. After our fit, values smaller than 1 imply that we have a better constraint on the PDF.
The hessian method provides a diagonalized covariance matrix of the parameters.
This fitting operation provides a new covariance matrix, $V$, in which values on the main diagonal are smaller than 1 and terms off diagonal can be different from zero. This matrix can be used to calculate uncertainties on PDFs using the usual equation for correlated variables:
$$\delta F = \sqrt{\sum_{ij}\frac{\partial F}{\partial \lambda_i} V_{ij}\frac{\partial F}{\partial \lambda_j}} $$
where $\lambda_i$ are the fitted parameters and $F$ is a PDF (e.g.e $u_V$, $d_V$ , $sea$ , $gluon$ etc.).

\subsection{PDFs built using a Neural network method}
NNPDF uses a neural network approach. PDFs are parametrized with a redundant flexible neural-net (i.e. a convenient functional form with $\sim 200$ parameters)
This technique produces a set of equally likely Montecarlo replicas of the PDFs. The best estimation of any observable of the PDFs is obtained by generating this observable using each of the replicas and averaging:\\ $\left< f\left[q\right]\right> = \frac{1}{N}\sum_{k=1}^{N}f\left[q^{(k)}\right]$  where $N$ is the number of replicas, $f$ is a generic function of the PDF set and $q^{(k)}$ are PDF replicas.
The uncertainty can be expressed through the standard deviation.
$\sigma_f^2=\left( \frac{1}{N-1}\sum_{k=1}^{N}\left(f\left[q^{(k)}\right] - \left< f\left[q\right]\right>
\right)^2  \right)$. If more than one observable is generated, it is even possible to compute correlations between quantities.
\\
To evaluate the PDF constraint with this PDFset we generated pseudo-experiment as average over the replicas then we generated rapidity distributions with each replica. We compared the prediction of a single replica with the pseudo-experiment minimizing the $\chi^2$
$$
\chi^2_i = \sum_{bin}\left(\frac{N_{bin}-\lambda_0f_i(y)}{\Delta_{bin}}\right)^2
$$
where  $i$ is the index of the replica,  $N_{bin}$ is the pseudo-experiments result in that rapidity-bin and $f_i(y)$ is the value of the rapidity distribution generated with the $i^{th}$ replica. This algorithm associates to each replica of the PDFs a $\chi^2$ and so the degree of compatibility of each replica can be evaluated as the chi-square probability. Now it is possible to constrain the PDFs considering the $\chi^2$-probability as the conditional probability for measuring the set of NNPDF replicas considering the new LHCb data \cite{bayes}. We don't consider all the PDF replicas as equally likely anymore but we use the  $\chi^2$-probability as a weight in the  central value and uncertainties calculations.
\section{Results and comments}
\begin{figure}
     \centering
     \subfigure[Improvement when fitting $W^+,~ W^-~Z$ rapidity distributions and a combined fit of the three]{\includegraphics[width=0.7\columnwidth]{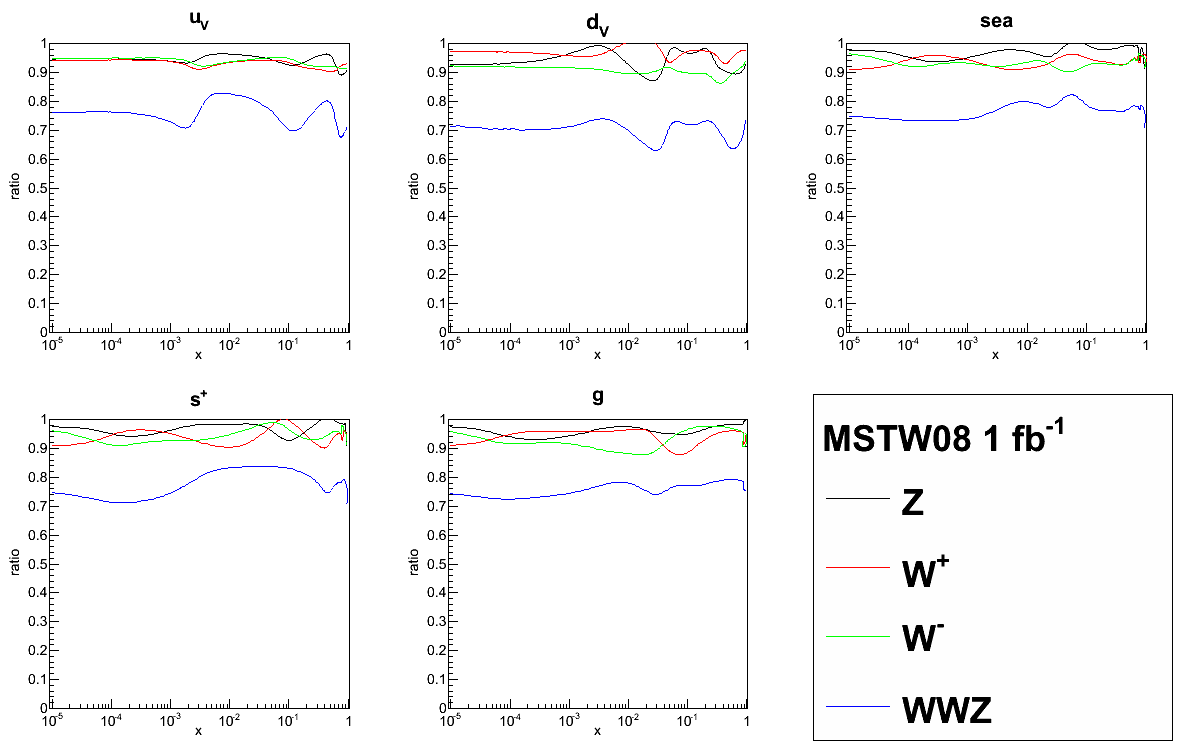}}
     \hspace{5mm}
     \subfigure[Improvement when fitting  Drell-Yan rapidity distributions in 4 different mass bins]{\includegraphics[width=0.7\columnwidth]{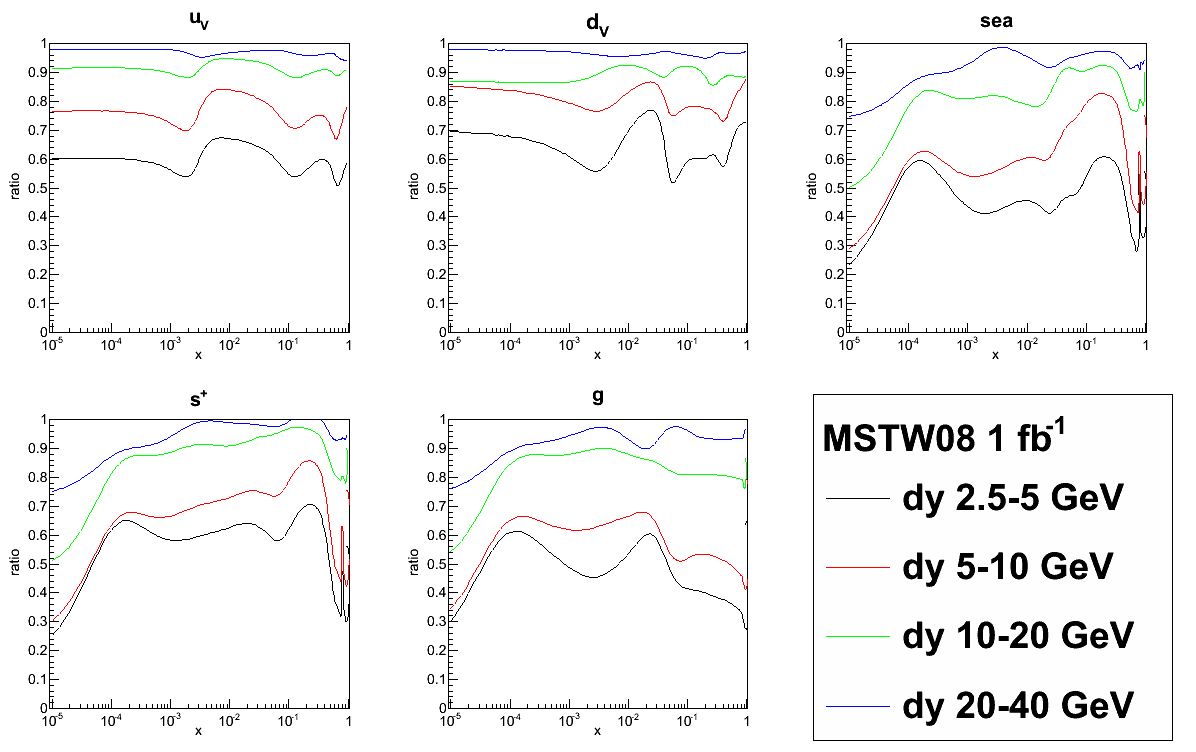}}
     \caption{Ratio between PDF uncertainties before and after adding $1~fb^{-1}$ of LHCb data for different distributions as a function of x}\label{res}
   \end{figure}

We present the expected improvement on \emph{u-valence, d-valence, sea , $s^+$  and gluon} distributions when introducing $1~fb^{-1}$ of LHCb electroweak data. Here we present  results obtained using the PDF set MSTW08. Results  for the other PDF sets can be found in \cite{myTalk}.
In figure \ref{res} the ratio between the PDF uncertainty after and before the fitting procedure is plotted. As expected the largest improvement is given by the lowest mass bin of the Drell-Yan process. With $1~fb^{-1}$ of LHCb data the $sea$ and the $gluon$ distributions at $x\sim10^{-5}$ and at high-x can be improved up to $70\%$ while in mid-x regions the improvement is about $50\%$. Other PDFs can be improved on the full $x$ range of about $60\%$. For the $W^{\pm}$ and $Z$ production the best result is obtained when performing a combined fit of the three rapidity distributions. About $30\%$ improvement can be expected over the whole $x$ range for each of the PDFs.


\begin{thebibliography}{99}
\bibitem{tdr} LHCb Collaboration,  "Reoptimised Detector Design and Performance " TDR, JINST3: 508005, 2008;
\bibitem{james} J. M. Keaveney,  "Studies of electroweak boson production in the forward region with LHCb"  in the same proceedings;
\bibitem{mstw} A.D. ~Martin, W.J. ~Stirling, R.S. ~Thorne, G. ~Watt,   "Parton distributions for the LHC", arXiv:hep-ex/0901.0002v2;
\bibitem{cteq} J. Pumplin, D.R. Stump, J. Huston, H.L. Lai, P. Nadolsky, W.K. Tung, "New Generation of Parton Distributions with Uncertainties from Global QCD Analysis", JHEP0207:012,2002;
\bibitem{nnpdf}R. D. Ball, L. Del Debbio, S. Forte, A. Guffanti, J. I. Latorre, A. Piccione, J. Rojo, M. Ubiali,"A first unbiased global NLO determination of parton distributions and their uncertainties" arxiv:hep/-ph:1002.4407;
\bibitem{alekh}S. I. Alekhin, Phys. Rev. {\bf D 63}, 094022 (2001); 
\bibitem{bayes} W. T. ~Giele, S. ~Keller "Implications of Hadron Collider Observables on Parton Distribution Function Uncertainties"  arXiv:hep-ph/9803393v1;
\bibitem{myTalk} F. De Lorenzi, Slides:\\
 \verb$http://indico.cern.ch/contributionDisplay.py?contribId=184&confId=86184$
\end{thebibliography}
\end{document}